\documentclass[preprint,amsmath,amssymb,prb,floatfix,superscriptaddress]{revtex4}

\usepackage[dvips]{graphicx}
\usepackage{dcolumn}
\usepackage{bm}
\usepackage{mathrsfs}

\begin{document}
\title{Terahertz wave generation via optical rectification from multiferroic BiFeO$_3$}
\author{D. Talbayev}
 \email{diyar@lanl.gov}
 \affiliation{Center for Integrated Nanotechnologies, MS K771, Los Alamos National Laboratory, Los Alamos, NM 87545, USA}
 \author{S. Lee}
\affiliation{Department of Physics, Rutgers University, 136 Frelinghuysen Rd., Piscataway, NJ 08854, USA}
\author{S.-W. Cheong} 
\affiliation{Department of Physics, Rutgers University, 136 Frelinghuysen Rd., Piscataway, NJ 08854, USA}
\author{A. J. Taylor} 
\affiliation{Center for Integrated Nanotechnologies, MS K771, Los Alamos National Laboratory, Los Alamos, NM 87545, USA}
\date{\today}

\newcommand{\cm}{\:\mathrm{cm}^{-1}}
\newcommand{\T}{\:\mathrm{T}}
\newcommand{\mc}{\:\mu\mathrm{m}}
\newcommand{\ve}{\varepsilon}
\newcommand{\dg}{^\mathtt{o}}

\begin{abstract}
We detected broadband coherent terahertz (THz) emission from multiferroic BiFeO$_3$ after illuminating a high-quality bulk single ferroelectric domain crystal with a $\sim$100 fs optical pulse. The dependence of the emitted THz waveform on the energy and polarization of the optical pulse is consistent with the optical rectification mechanism of THz emission. The  THz emission provides a sensitive probe of the electric polarization state of BiFeO$_3$, enabling applications in ferroelectric memories and ferroelectric domain imaging. We also report room-temperature THz optical constants of BiFeO$_3$.
\end{abstract}

\maketitle

Multiferroic BiFeO$_3$ (BFO) offers the possibility of electrical manipulation of its magnetic state, which can be used in devices such as electrically-controlled magnetic memory elements\cite{zhao:823,ederer:60401}. At room temperature, single-crystalline BFO is found in a G-type antiferromagnetic state ($T_N$=640 K) with a modulated cycloidal spin structure with a long modulation period of $\sim$620 \AA\cite{sosnowska:4835, lee:192906, lee:100101, lebeugle:227602}. The material is also a robust ferroelectric with a high Curie temperature $T_C$ of 1,103 K and large ferroelectric polarization of $\sim$100  $\mu$C/cm$^2$\cite{lebeugle:22907}. Nonlinear optics plays an important role in characterization of ferroelectrics, as the lack of inversion symmetry in ferroelectric crystals results in the presence of strong second-order nonlinearities; nonlinear optics is also an important area for ferroelectric applications, along with transducers and data storage. Linear and nonlinear optical properties of BFO have been reported\cite{lobo:172105, fruth:937, kumar:121915, ihlefeld:142908}. Recently, it has been proposed that terahertz (THz) radiation from BFO thin films can be used in ferroelectric domain imaging and optical readout of the state of ferroelectric memories\cite{takahashi:117402, takahashi:52908, rana:24105}. In that work, the THz wave was generated by ulrafast modulation of ferroelectric polarization when charge carriers were injected in the film by a femtosecond laser pulse.

In this Letter, we report coherent generation of THz waveforms by femtosecond 800-nm optical pulses from ferroelectric BFO in reflection and transmission geometries and also provide the THz-frequency optical constants of BFO. The THz wave is generated via optical rectification that relies on the second-order nonlinearity. The amplitude of the THz wave detected in the time-domain depends on the direction of the ferroelectric dipole in the BFO crystal. In contrast, a measurement of the intensity of the second-harmonic light does not provide this information\cite{kumar:121915, lofland:92408}. Thus, THz emission provides a sensitive probe of the polarization state of the BFO crystals. The presented work was performed on a high-quality single ferroelectric domain crystal of BFO, which is evidenced by the pronounced polarization dependence of both the optical rectification and the measured THz optical constants. Such high-quality crystals are indispensable for the reliable characterization of the basic material properties\cite{lebeugle:227602,mathur:591}.

For this work, bulk single ferroelectric domain BFO crystal was grown using a Bi$_2$O$_3$ flux. The ferroelectric phase possesses the rhombohedral $R3c$ structure that can be obtained by a slight elongation of the pseudocubic perovskite unit cell along the (111) direction. Our measurements were performed at room temperature on a 0.22 mm thick (001)$_{cubic}$-oriented crystal mounted on a 3-mm diameter aperture. Coherent THz transients were generated upon illumination of the crystal by 800-nm femtosecond pump pulses provided by Ti:Sapphire regenerative amplifier operating at 1 kHz and focused to a 2-mm diameter spot. The maximum pump power of 150 mW used in these measurements corresponds to the fluence of 4.8 mJ/cm$^2$. The THz transients emitted from BFO were focused on a [110]-oriented ZnTe crystal for coherent detection via electro-optic sampling\cite{wu:3523, nahata:2321}. The THz optical constants of the BFO crystal were measured in transmission in a time-domain spectroscopy system that uses photoconductive antennas for THz generation and detection\cite{vanexter:337}.

\begin{figure}[ht]
\begin{center}
\includegraphics[width=3in]{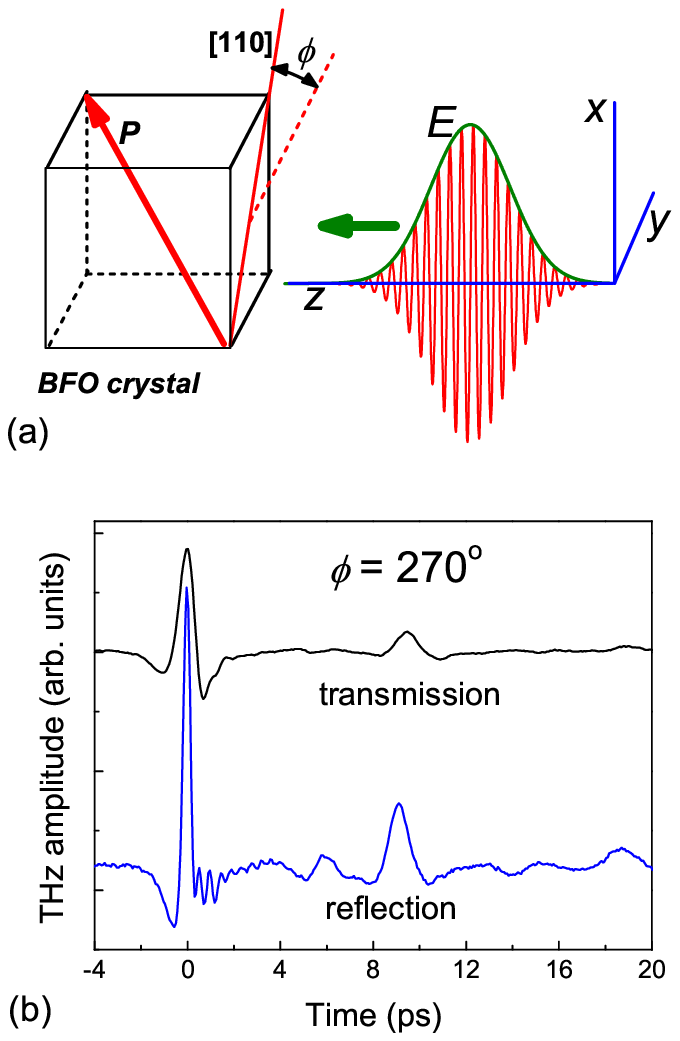}
\caption{\label{fig:thzwave}(Color online) (a) Geometry of THz generation in (001)$_{cubic}$-oriented BFO. The pump light is polarized in the $x$-$z$ plane. Static ferroelectric polarization $P$ is along the [111] direction (the diagonal of the cube). (b) THz waveforms detected in transmission and reflection geometry, pump fluence 4.8 mJ/cm$^2$. The main THz pulse and the first Fabry-P\'erot reflection are shown.}
\end{center}
\end{figure}

Figure~\ref{fig:thzwave} shows the geometry of THz generation measurement in BFO. The incident femtosecond pump pulses are polarized in the horizontal $x$-$z$ plane. The ferroelectric dipole moment in BFO points along the long diagonal of the distorted perovskite structure and makes a $35\dg$ angle with the surface of our crystal. Figure~\ref{fig:thzwave} displays the THz transients emitted from the crystal in transmission and reflection geometries. Both in transmission and reflection, the main THz pulse is followed by the Fabry-P\'erot (or etalon) reflection within the sample.

\begin{figure}[ht]
\begin{center}
\includegraphics[width=3in]{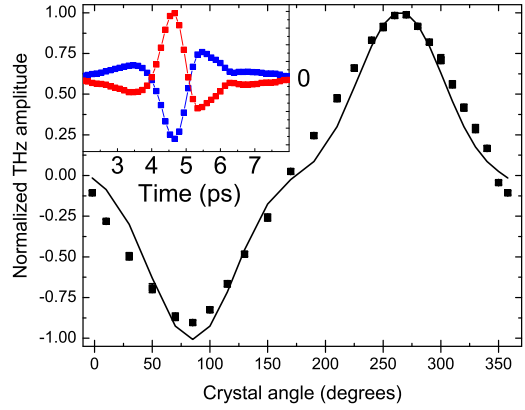}
\caption{\label{fig:ampangle}(Color online) Amplitude of the main THz pulse as a function of the angle $\phi$ between the $y$ axis and the [110] direction in the crystal while rotating the crystal about the $z$ axis. The solid line shows the simulated nonlinear polarization $P(t)$. The inset shows the main THz pulse at $\phi=270\dg$ and $\phi=90\dg$, the positions of maximum positive and negative THz amplitudes.}
\end{center}
\end{figure}

\begin{figure}[ht]
\begin{center}
\includegraphics[width=3in]{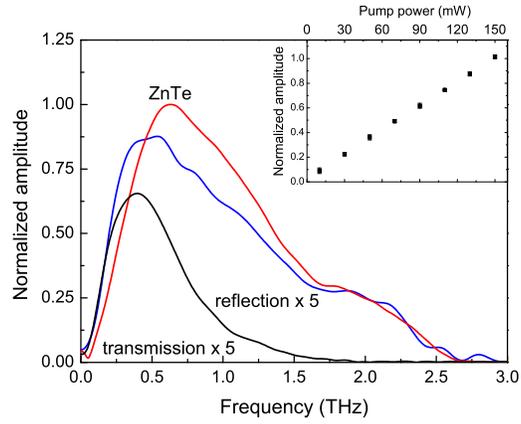}
\caption{\label{fig:fdsp}(Color online) Frequency-domain spectra of the waveforms emitted from BiFeO$_3$ and from the [110] oriented ZnTe crystal. The THz waveforms were measured with the pump power of 150 mW and 28 mW, respectively. The inset shows the amplitude of the THz wave from BiFeO$_3$ detected in transmission as a function of pump power.}
\end{center}
\end{figure}

When a lightwave of frequency $\omega$ is incident on a crystal with a second-order optical nonlinearity $\chi^{(2)}$, the polarization induced by the lightwave contains the nonlinear part $P^{NL}=\chi^{(2)}(\cos\omega t)^2 = 1/2\chi^{(2)}(1 + \cos2\omega) = P^{NL}(0) + P^{NL}(2\omega)$. The latter term in the last expression describes the second harmonic generation. The former term describes a zero-frequency polarization, which is truly a constant (DC) polarization induced in the crystal when the incident lightwave is a sinusoidal field of infinite duration. Induction of DC polarization due to the second-order nonlinearity is known as optical rectification. When the incident lightwave is a 100-fs-long pulse (Fig.~\ref{fig:thzwave}a), the induced $P^{NL}(0)$ only exists for the duration of the pulse and represents a time-varying electric dipole $P(t)$. This time-dependent dipole emits a THz electromagnetic wave $E_{THz}(t)\sim \partial^2P(t)/\partial t^2$. To confirm that optical rectification is responsible for the observed THz emission, we measure the dependence of the emitted THz amplitude on the angle between the [110]$_{cubic}$ crystallographic direction and the $y$ axis while rotating the crystal about its normal (Fig.~\ref{fig:thzwave}a). The angular dependence of the emitted THz field is consistent with the optical rectification mechanism. We model the induced second order THz-frequency polarization $P_{i} = \chi^{(2)}_{ijk}E_j(\omega)E_k(\omega)$ using the set of nonlinear optical constants of BFO films found by Kumar $et$ $al$\cite{kumar:121915}. The emitted THz field $E_{THz}(t)$ has the same angular dependence as the induced $P(t)$. The solid line in Fig.~\ref{fig:ampangle} represents the calculated induced polarization and shows excellent agreement with the measured THz field amplitude. The meaning of the negative THz amplitude is illustrated in the inset of Fig~\ref{fig:ampangle}. The optical rectification mechanism of the THz emission is supported further by the linear dependence of the THz field amplitude on the pump power (inset of Fig.~\ref{fig:fdsp}).

The ferroelectric axis of BFO can be along any of the four long diagonals of the pseudocubic unit cell (Fig.~\ref{fig:thzwave}a). Correspondingly, eight different ferroelectric domains are possible in the crystal. The pronounced angular dependence of THz emission (Fig.~\ref{fig:ampangle}) demonstrates that our crystal predominantly consists of a single ferroelectric domain, which is a result of crystal growth below the Curie temperature. The single-domain nature of these millimeter-sized crystals was confirmed independently by neutron diffraction\cite{lee:192906, lee:100101}.

A different mechanism for THz emission from BFO thin films was described by Takahashi $et$ $al$\cite{takahashi:117402, takahashi:52908, rana:24105}. In their work, the emission is due to the injection of photoexcited charge carriers and ultrafast modulation of spontaneous polarization $P_f$. In our measurements, similar processes involving the injection of photocarriers can be ruled out because the energy of our 800 nm, 1.5 eV pump pulses is lower than the direct optical gap of BFO. To confirm this, we carried out optical-pump THz-probe study of the BFO crystal, in which no changes to the THz optical constants were induced by the 800-nm pump pulses. The absence of THz conductivity modulation indicates that no free carriers are created by the 800-nm light\cite{averitt:1357}. In contrast, Takahashi $et$ $al$ used 400-nm pump light in their work to inject the photocarriers.  Our BFO crystal is opaque to the 800-nm light, even though this photon energy is lower than the measured bandgap. The residual absorption is most likely due to a certain level of present defects. The same defects are expected to act as traps for the few carries that are created by the absorption. Thus, the residual absorption in our BFO crystal does not result in a measurable change in THz conductivity.

We determine the spectral content of THz pulses emitted from BFO by taking a Fourier transform of the measured time-domain transients and compare it to THz emission from 0.5-mm thick [110]-oriented ZnTe crystal (Fig.~\ref{fig:fdsp}). The THz spectrum emitted from BFO in reflection is very similar to that of the emission from ZnTe, although the amplitude of ZnTe emission is approximately 25 times larger than that of BFO at the same pump power. The differing emission conditions in ZnTe and BFO, i.e., different crystal thicknesses, emission direction and phase matching, suggest that in both cases the  $\sim 2.5$ THz bandwidth may be limited by the frequency response of the electro-optic-sampling THz detector used in the measurement.

\begin{figure}[ht]
\begin{center}
\includegraphics[width=3in]{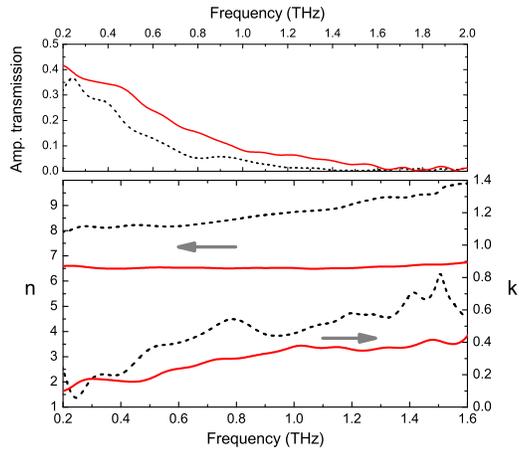}
\caption{\label{fig:trspnk}(Color online) Upper panel: Amplitude transmission spectra of BiFeO$_3$ crystal for THz light polarized in the plane of the electric dipole (solid line) and perpendicular to the plane (dotted line). Lower panel: real and imaginary parts of the  refractive index for the polarizations displayed in the upper panel.}
\end{center}
\end{figure}

The spectrum of the THz pulse emitted from BFO in transmission is considerably narrower than the spectrum of the pulse emitted in reflection (Fig.~\ref{fig:fdsp}; in this Figure all spectra are normalized by the maximum amplitude of the ZnTe spectrum). We attribute this to the absorption of the higher frequencies in the THz pulse by BFO as the pulse propagates through the crystal. We point out that even though the optical absorption at 800 nm is negligible compared to the above-bandgap absorption in BFO, the residual absorption is sufficiently strong to prevent the 800-nm light from propagating far into the BFO crystal. We estimate that the penetration depth of the 800-nm light is about $10\mc$, which is only a small fraction of the total $220\mc$ thickness of our BFO crystal. (Our BFO crystal is completely opaque to the eye. A different BFO crystal became transparent after we polished it down to $10\mc$ thickness.) Therefore, the THz wave is emitted in a thin surface layer of our BFO crystal and then propagates through and is absorbed by the remaining portion of the crystal. To confirm this, we measured THz absorption in BFO (upper panel of Fig.~\ref{fig:trspnk}) and found that the crystal is indeed opaque above $\sim 1.6$ THz for the light polarized along the ferroelectric polarization $P_f$. The THz optical constants extracted from the measurement (lower panel of Fig.~\ref{fig:trspnk}) allow an estimate of the THz pulse amplitude emitted from the $10$-$\mc$-thick layer in which the pump pulse is absorbed. Thus calculated THz amplitude is comparable to the measured amplitude of the THz pulse emitted in reflection. The two amplitudes are expected to be the same because phase matching is unimportant in a sufficiently thin crystal. More evidence for THz emission occurring in a thin surface layer of the crystal is provided by the broadened Fabry-P\'erot reflection of the main THz pulse detected in reflection (Fig.~\ref{fig:thzwave}). The time-domain width of the Fabry-P\'erot pulse is the same as that of the main transmitted THz pulse, which means that the main reflected pulse is narrow in the time domain and has a much broader spectrum in the frequency domain because it does not traverse an appreciable thickness of BFO material and therefore does not lose the high-frequency components in its spectrum.

In conclusion, we demonstrate the emission of coherent THz radiation from BFO due to optical rectification of $\sim100$ fs 800-nm pump pulses both in reflection and transmission geometries. This mechanism of THz emission is different from the previously reported emission via ultrafast modulation of the spontaneous ferroelectric polarization\cite{takahashi:117402, takahashi:52908, rana:24105}. The potential applications of coherent THz emission include ferroelectric domain imaging and non-destructive readout of the state of ferroelectric memory elements. Since optical rectification does not require the absorption of the pump pulse, our demonstration opens the door for future studies using pump wavelengths longer that the 442-nm bandgap of BFO, including the important $1.3\mc$ and $1.55\mc$ telecommunications wavelengths. We also report room-temperature THz-frequency optical constants of BFO, which will be indispensable in modeling THz emission and propagation in those devices.

We would like to thank Kiyong Kim for useful discussions. The work at LANL was supported by the LDRD program and by the Center for Integrated Nanotechnologies. The work at Rutgers was supported by NSF-DMR-0520471.

\newpage
\renewcommand{\baselinestretch}{2}

\newpage

\newpage

\newpage

\newpage
\end{document}